\documentstyle[12pt,aasms4]{article}
\newcommand{\postscript}[2]
 {\setlength{\epsfxsize}{#2\hsize}
  \centerline{\epsfbox{#1}}}
\def\tempest%
{\begin{array}{ccc}
1 & 1 & 1 \\
1 & 1 & 1 \\
4 & 3 & 8
\end{array}}

\begin{document}

\title{Inventory of Gravitational Microlenses\\
	Toward the Galactic Bulge}
\author{Cheongho Han}
\author{Seunghun Lee}
\bigskip
\affil{Dept.\ of Astronomy \& Space Science, \\
       Chungbuk National University, Cheongju, Korea 361-763 \\
       cheongho@astro.chungbuk.ac.kr,\  leess@astro.chungbuk.ac.kr}
\authoremail{cheongho@astro.chungbuk.ac.kr}
\authoremail{leess@astro.chungbuk.ac.kr}

\begin{abstract}
We determine the microlensing event rate distribution, 
$\Gamma(t_{\rm E})$, that is properly normalized by the 
first year MACHO group observations (Alcock et al.\ 1997).\markcite{alcockb}
By comparing the determined $\Gamma(t_{\rm E})$ with various 
MF models of lens populations to the observed distribution, we find that 
stars and WDs explain just $\sim 49\%$ of the total event rate 
and $\sim 37\%$ of the observationally determined optical depth
even including very faint stars just above hydrogen-burning limit.
Additionally, the expected time scale distribution of events  caused by 
these known populations of lenses deviates significantly
from the observed distribution especially in short time scale region.
However, if the rest of the dynamically measured mass of the bulge 
($\sim 2.1\times 10^{10}\ M_\odot$) is composed of brown dwarfs, 
both expected and observed event rate distribution matches very well.
\end{abstract}
 
\vskip100mm

\bigskip
\bigskip
\centerline{submitted to {\it The Astrophysical Journal}: May 19, 1997}
\centerline{Preprint: CNU-A\&SS-02/97}
\clearpage

\section{Introduction}
The gravitational microlensing experiments toward Large Magellanic Cloud 
(Ansari et al.\ 1996; Alcock et al.\ 1996)\markcite{ansari}\markcite{alcocka}
was originally proposed to search for Massive Astronomical Compact 
Halo Objects (MACHOs) (Paczy\'nski 1986; Griest 1991).
\markcite{paczynskia}\markcite{griest} 
On the other hand, the experiments toward the Galactic bulge 
(Alcock et al.\ 1997; Udalski et al.\ 1994; Alard 1996) 
\markcite{alcockb}\markcite{udalski1994}\markcite{alard}
were initiated mainly to check the feasibility of the experiments 
since a certain event rate is expected from the ``known'' population 
of lenses, i.e., stars.
However, experiments toward the galactic bulge not only provide
a new probe of the galactic structure but also become
an important tool that can resolve the controversial 
``disk dark matter'' problem (Bahcall 1986; Kuijken \& Gilmore 1989).
\markcite{bahcall}\markcite{kuijken}

There already have been several trials to determine the mass function (MF)
of galactic lenses from the result of microlensing experiments
(Zhao, Spergel, \& Rich 1995;  Zhao, Rich, \& Spergel 1996; 
Jetzer 1994; Han \& Gould 1996).
\markcite{zhaoa}\markcite{zhaob}\markcite{jetzer}\markcite{hanb}
However, these analyses either determined event rate distribution 
by assuming some arbitrary functional forms of MF, 
e.g., power law, or could provide at most very crude information 
about lenses.
For poorly-known lens populations such as brown dwarfs (BDs), 
this type of analysis might be the only approach currently possible, 
and thus the derived MF based on the observed Einstein time scale 
distribution is very uncertain and model dependent 
(Mao \& Paczy\'nski 1996; Gould 1996). 
\markcite{Mao}\markcite{gouldb}
The Einstein time scale is related to the physical parameters of 
a lens by 
$$
t_{\rm E} = {r_{\rm E} \over v},\qquad
r_{\rm E} = \left( {4Gm \over c^2}
{D_{\rm ol}D_{\rm ls}\over D_{\rm os}}\right)^{1/2},
\eqno(1.1)
$$
where $r_{\rm E}$ is the Einstein ring radius, $v$ is the source-lens
transverse speed, $m$ is the mass of the lens, and $D_{\rm ol}$,
$D_{\rm ls}$, and $D_{\rm os}$ are the separations between the 
observer, lens, and source.
The better approach will be, however, first estimate the expected 
event rate distribution from the well constrained lens populations, 
e.g., stars and white dwarfs (WDs), and then 
test other possible lens populations.
By doing this, one can set the upper limits on dark lens populations. 
However, still no quantative estimate of detailed event rate and 
time scale distribution by even known population of lenses 
has been made.

In this paper, we determine the microlensing event rate distribution, 
$\Gamma(t_{\rm E})$, that is properly normalized by the 
first year MACHO group observations (Alcock et al.\ 1997).\markcite{alcockb}
By comparing the determined $\Gamma(t_{\rm E})$ with various 
MF models of lens populations to the observed distribution, we find that 
stars and WDs explain just $\sim 49\%$ of the total event rate 
and $\sim 37\%$ of the observationally determined optical depth
even including very faint stars just above hydrogen-burning limit.
Additionally, the expected time scale distribution of events  caused by 
these known populations of lenses deviates significantly
from the observed distribution especially in short time scale region.
However, if the rest of the dynamically measured mass of the bulge 
($\sim 2.1\times 10^{10}\ M_\odot$, Zhao, Spergel, \& Rich) is 
composed of brown dwarfs (BDs), both expected and observed event 
rate distribution matches very well.

\section{Model I: Lenses of Known Population}

\subsection{Mass Function}

To determine the microlensing event rate distribution 
caused by stars, it is required to construct 
properly normalized MF of galactic stars.
Unfortunately, the MF cannot be obtained from observation, 
because it is difficult to measure stellar masses.
Instead, one can construct MF by inferring stellar masses from well 
constrained luminosity function (LF) by using mass-luminosity relation.

The LF of galactic bulge stars is constructed as follows.
For stars brighter than $M_V\sim 4$, most of these stars are 
clump giant stars, the LF is constructed based on the de-reddened $I$-band 
LF determined by J. Frogel (private communication).
For the part of the LF fainter than $M_V \sim 4$, we adopt that 
determined by Light, Baum, \& Holtzmann (1997)\markcite{light}
by using the Hubble Space Telescope.
They find that the LF of stars in an angular size of ${\rm arcmin}^2$
is well represented by
$$
\log \Phi_V = 0.083 M_V + 2.47,
\eqno(2.1.1)
$$ 
in the magnitude range $4\leq M_V \leq 10$.
For stars even fainter than $M_V = 10$,
we extend the LF by adopting that of stars in the solar neighborhood 
(Gould, Bahcall, \& Flynn 1996) \markcite{goulda} under the assumption that 
galactic disk and bulge stars have similar LFs.
The absolute $V$ band magnitude is determined based on a distance 
modulus of $\mu_{\rm bulge}=14.5$, which is equivalent to the distance of 
8 kpc to the bulge.
For the luminosity to mass conversion we use 
the mass-luminosity relation determined by
Henry \& McCarthy (1993).\markcite{henry} 
The relation has the form
$$
\log(m/M_{\odot}) =
\cases{
0.4365-0.0971M_V + 0.002456M_{V}^2,   & ($M_V \le 10.25$), \cr
1.4217 - 0.1681M_V,                   & ($10.25 \le M_V \le 12.89$), \cr
1.4124 - 0.2351M_V + 0.005257M_{V}^2, & ($M_V \ge 12.89$). \cr
}
\eqno(2.1.2)
$$
Since the mass-luminosity relation uses $V$-band magnitude, we convert the 
$I$-band to $V$-band magnitude by
$$
M_V = \left( {3.37 M_I - 2.89 \over 2.37}\right).
\eqno(2.1.3)
$$

In addition to visible stars, we also include WDs
into a member of known lens populations 
(Adams \& Laughlin 1996).\markcite{adamsb} 
In our model MF, WDs are uniformly distributed in the
mass range $0.5\ M_\odot \leq m \leq 0.6\ M_\odot$ and their total number 
is normalized so that there are 10 times more WDs than clump giants
(stars brighter than $M_V \lesssim 4$) considering the life expectancy of 
clump giants and age of the Universe.

The model LF, $\Phi_L dM_I$, and the corresponding MF, $\Phi_m dm$, 
are shown in Figure 1 and in the upper panel of Figure 2, respectively.
Both functions are normalized for objects in a physical area of 
${\rm pc}^2$ at the galactic center, corresponding to an angular area of 
$(0.43\ {\rm arcmin})^2$.
In the figure, 
the range of LF, $-4 \leq M_I \leq 15$, is divided into 1000 intervals,
while that of the MF, $0\ M_\odot \leq m \leq 1.5\ M_\odot$, is
divided into 100 intervals.

Then what fraction the galactic bulge mass is composed of 
these stellar and WD lenses?
These values are determined from the relation
$$
{\Sigma_L \over L_{\rm bulge}} =
{\Sigma_m \over m_{\rm pop}},
\eqno(2.1.4)
$$
where $L_{\rm bulge} = \int_{\rm bulge} \nu dV$ is the total luminosity
of the bulge and $\Sigma_L = \int_{\rm BW} \nu d\ell$ is the surface 
light density seen through BW.
Here $\nu$ is the 3-dimensional light density distribution of the
galactic bulge and the notations $\int_{\rm bulge} \nu dV$
and $\int_{\rm BW} \nu d\ell$ represent volume integral over the whole 
bulge and line integral along the line of sight toward BW, respectively.
For the  computation of $m_{\rm pop}$, we adopt Kent (1992)\markcite{kent}
bulge light density distribution model of the form
$$
\nu(s) =
\cases{
1.04\times 10^6\ (s/0.482)^{-1.85}\  L_{\odot}\ {\rm pc}^{-3},
&  $(s < 938\ {\rm pc})$, \cr
3.53 K_0\ (s/667)\ L_{\odot}\ {\rm pc}^{-3},
& $(s \ge 938\ {\rm pc})$, \cr
}
\eqno(2.1.5)
$$
where $s^4 = R^4 +(z/0.61)^4$, the coordinates $(R,z)$ represent
the galactocentric distance along and normal to the galactic plane,
and $K_0$ is a modified Bessel function.
From this light density model, we find the surface light density of 
$\Sigma_L = 2412\ L_\odot\ {\rm pc}^{-2}$ and 
the total luminosity of
$L_{\rm bulge} =  1.8\times 10^{10}\ L_\odot$. 
In addition, the surface number and
mass densities of individual lens populations toward Baade's 
Window (BW) are computed from the MF by 
$$
\Sigma_N= \int_{\rm BW} \Phi_m dm, 
$$
$$
\Sigma_m= \int_{\rm BW} m\Phi_m dm, 
\eqno(2.1.6)
$$
resulting in 
$\Sigma_N=3422\ {\rm pc}^{-2}$ and 
$\Sigma_m=1562\ M_\odot\ {\rm pc}^{-2}$
for stars, and 
$\Sigma_N=614\ {\rm  pc}^{-2}$ and $\Sigma_m=369\ {\rm  pc}^{-2}$ for WDs.
Then the total masses of individual populations in the bulge are
$m_{\rm star}=1.18\times 10^{10}\ M_\odot$ and
$m_{\rm WD}=0.25\times 10^{10}\ M_\odot$ for stars and WDs, respectively.
According to this MF model,
the combined total mass of stars and WDs in the bulge  
comprises $\sim 70\%$ of the dynamically determined bulge mass of 
$2.1\times 10^{10}\ M_\odot$ (Zhao, Spergel, \& Rich 1995).\markcite{zhaoa}
In Figure 1, we list the values of $\Sigma_N$, $\Sigma_m$, and $\Sigma_L$,
and resultant $m_{\rm pop}$.

\subsection{Event Rate}

The next question is, then, how many events are caused by these 
known populations of lenses?
The event rate distribution of bulge self-lensing events
(Paczy\'nsky et al.\ 1994) \markcite{paczynskib} for a single source 
is computed by
$$
\Gamma'_{\rm bulge}(t_{\rm E}) = 
\epsilon(t_{\rm E})
\left[\int dD_{\rm os} n(D_{\rm os}) 
\int dD_{\rm ol}\ (2 r_{\rm E}) n(D_{\rm ol})
\int dv_y dv_z vf(v_y,v_z)
\ \delta ( t_{\rm E}-t'_{\rm E}) \right]
$$
$$
\times
\left[ \int dD_{\rm os} n(D_{\rm os})\right]^{-1},
\eqno(2.2.1)
$$
where $\epsilon (t_{\rm E})$ is the detection efficiency,
$n(D_{\rm os})$ and $n(D_{\rm ol})$ are the number densities
of source stars and lenses, $(v_{y},v_{z})$ are the components of 
the transverse velocity, ${\bf v}$, parallel and normal to 
the galactic plane,  and $f(v_y,v_z)$ represents their distributions.
The factor $2r_{\rm E}$, i.e., Einstein ring diameter, and $v$ are 
included because lenses with larger cross-sections and higher 
transverse speeds result in higher event rate.
For the case of disk-bulge events, on the other hand, the rate 
distribution computation can be simplfied by 
$$
\Gamma'_{\rm disk} (t_{\rm E})= 
\epsilon(t_{\rm E})
\int dD_{\rm ol}\ (2r_{\rm E}) n(D_{\rm ol}) \int dv_y dv_z vf(v_y,v_z)
\ \delta ( t_{\rm E}-t'_{\rm E}).
\eqno(2.2.2)
$$
because bulge stars are located at a large distance, i.e., 
$D_{\rm os}\sim 8\ {\rm kpc}$, compared to a typical
observer-lens separation. 

For the galactic bulge and disk matter distributions, we adopt Kent
bulge model (see eq.\ [2.1.6]) and double exponential disk model.
The disk matter distribution model has the form
$$
n(R,z) = n_0 \exp \left[ - \left( {R-8000 \over h_R}  
+ {z \over h_z} \right) \right],
\eqno(2.2.3)
$$
where the values of the radial and vertical scale heights are
$h_R = 3.5\ {\rm kpc}$ and $h_z = 325\ {\rm pc}$, respectively,
and $n_0$ is the normalization (see below).
The velocity distributions for both disk and bulge components
are modeled by a gaussian, i.e.,
$$
f(v_i)\propto \exp \left[ -{(v_i-\bar{v}_i)\over 2\sigma_i^2}\right];
\qquad
i \in {y,z},
\eqno(2.2.4)
$$
where the means and dispersions are 
$(\bar{v}_{y,{\rm disk}},\sigma_{y,{\rm disk}})=(220,30)\ {\rm km\ s}^{-1}$
and 
$(\bar{v}_{z,{\rm disk}},\sigma_{z,{\rm disk}})=(0,20)\ {\rm km\ s}^{-1}$
for the disk component, and
$(\bar{v}_{y,{\rm bulge}},\sigma_{y,{\rm bulge}})=(0,93)\ {\rm km\ s}^{-1}$
and 
$(\bar{v}_{z,{\rm bulge}},\sigma_{z,{\rm bulge}})=(0,79)\ {\rm km\ s}^{-1}$
for the bulge component, respectively.
See Han \& Gould (1995, 1996)\markcite{hana,hanb} for more details.

For the computation of event rate distribution it is required to
know the 3-dimensional number density $n$, but what is observationally
known for the bulge matter distribution is the light density $\nu $.
Under the assumption that light is distributed the same way as
matter is distributed, one can convert light density into number 
density by multiplying the number-to-light ratio, i.e.,
$$
n(D_{\rm ol}) = \left( {N\over L} \right)\nu(D_{\rm ol}).
\eqno(2.2.5)
$$
We find the number-to-light ratio of the galactic bulge to be  
$N/L = \Sigma_N /\Sigma_L = 1.66$.
After normalizing the bulge self-lensing event rate distribution 
by using $N/L$ ratio, 
the disk-bulge event rate is scaled so that the optical depth
ratio between the two components becomes
$$
{\tau_{\rm disk} \over \tau_{\rm bulge}} = 
{
\int [ \Gamma'_{\rm disk} (t_{\rm E}) t_{\rm E}/
\epsilon (t_{\rm E})] dt_{\rm E} 
\over
\int [ \Gamma'_{\rm bulge} (t_{\rm E}) t_{\rm E} /
\epsilon (t_{\rm E})] dt_{\rm E}} =
{0.5 \times 10^{-6} \over 1.2 \times 10^{-6}},
\eqno(2.2.6)
$$
based on the optical depth computation by 
Han \& Gould (1995).\markcite{hana}

Once the event rate distribution for a single star is obtained by
equations (2.2.1) and (2.2.2), the total event rate distribution 
for all monitored $N_\ast$ stars during the total monitored time $T$ 
is computed by
$$
\Gamma (t_{\rm E}) = N_\ast T\Gamma' (t_{\rm E});
\qquad \Gamma' = \Gamma'_{\rm bulge} + \Gamma'_{\rm disk}.
\eqno(2.2.7)
$$
We find the total exposure of MACHO experiment by 
$$
N_\ast T = \left( {\pi \over 2\tau }\right) 
\sum_{i=1}^{N_{\rm event}} 
{ t_{{\rm E},i}\over \epsilon (t_{{\rm E},i}) }.
\eqno(2.2.8)
$$
MACHO group reported 45 events toward the galactic bulge 
(Alcock et al.\ 1997).\markcite{alcockb}
They determined the optical depth based on 41 events, excluding 
4 events either failed the final cut or suspected as variable stars, 
to be $\tau = 2.4\times 10^{-6}$.
By using the detection efficiency $\epsilon (t_{\rm E})$ also provided by 
MACHO group, we find the total exposure to be
$$
N_\ast T =  2.08\times 10^9\ {\rm stars}\ {\rm days}.
\eqno(2.2.9)
$$

The finally determined event rate distribution expected
from the known lens population (thick solid line) is shown
in the lower panel of Figure 2, and it is compared with the
observed distribution (histogram).
In the figure, the bulge-bulge and disk-bulge event rate distributions
are represented by short and long dashed lines, respectively. 
One finds that the event rate expected from the known populations of 
stars and WDs, $\Gamma_{\rm exp}$, alone cannot 
explain the observed event rate distribution, $\Gamma_{\rm obs}$.
One also finds that 
the biggest difference between the two rate distributions arises 
in short time scale region.
We find that events caused by stellar and WD lenses make up only
$\Gamma_{\rm exp,tot}/\Gamma_{\rm obs,tot} = 
\int \Gamma_{\rm exp} (t_{\rm E})dt_{\rm E}
/\int \Gamma_{\rm obs} (t_{\rm E}) dt_{\rm E} = 49\%$
of the total observed event rate, and they comprises 
$\tau_{\rm exp}/\tau_{\rm obs} = 
\int \left[ t_{\rm E}\Gamma_{\rm exp} (t_{\rm E})/\epsilon (t_{\rm E})
\right] dt_{\rm E}
/\int [t_{\rm E}\Gamma_{\rm obs} (t_{\rm E}) /\epsilon (t_{\rm E})a]
dt_{\rm E} = 37\%$
of the optical depth.

\section{Alternative Mass Function Models}

\subsection{Model II: Additional Brown Dwarf Population Lenses}

In previous section, we show that additional population(s) of 
lenses are required to explain the observed galactic bulge event rate 
distribution.
These candidate lens populations might be black holes, 
neutron stars (Venkatesan, Olinto, \& Truran 1997),\markcite{venkatesan} 
BDs, etc. 
Among these candidates, the most probable 
population would be BDs because the major difference 
between $\Gamma_{\rm obs}$ and $\Gamma_{\rm exp}$ arises in short 
time scale region.

Therefore, we make an alternative MF model in which 
lenses are composed of already known populations of stars and WDs plus
BDs which make up the rest of the dynamically determined galactic bulge 
mass.
For this case, BDs comprise 31\% of the total galactic bulge mass, 
corresponding to $0.65\times 10^{10}\ M_\odot$.
This fraction of BDs in the Galaxy might sound too big according to 
the theory of star formation (Adams \& Fatuzzo 1996; Graff \& 
Freese 1996).\markcite{adamsa}\markcite{graff}
However, one cannot rule out the possibility that there are a large 
number of BDs from a different population of stars, e.g., Population III.
In the MF model we assume BDs are uniformly distributed in the 
mass range $0.06\ M_\odot \leq m \leq 0.08\ M_\odot$.

The event rate with this alternative MF model is determined
similarly, but with different value of number-to-light ratio.
This is because the surface number density increases compared to that of 
model I, while the surface light density does not change (see Table 1)
due to the dark nature of BDs.
In Figure 3, we present the model MF and corresponding event rate 
distribution.
One finds that the determined event rate distribution including BD 
population matches very well with the observed distribution.
The ratio between expected and observed event rates and optical depths 
are $\Gamma_{\rm exp,tot}/\Gamma_{\rm obs,tot} = 0.96$ and
$\tau_{\rm exp}/\tau_{\rm obs} = 0.54$, respectively.
There are three very long time scale events with 
$t_{\rm E}\gtrsim 70\ {\rm days}$ whose nature is hard to understand
under reasonable matter and velocity distribution models.
If these three very long events are not included, the ratios become 
$\Gamma_{\rm exp,tot}/\Gamma_{\rm obs,tot} = 1.03$ and
$\tau_{\rm exp}/\tau_{\rm obs} = 1.28$.

\subsection{Model III: Power-law Mass Function}

Zhao, Spergel, \& Rich (1995)\markcite{zhaoa} claimed that the observed 
time scale distribution could be explained if the dynamically measured 
mass of the bulge were distributed in a Salpeter power-law MF between 
$0.08\ M_\odot$ and $0.6\ M_\odot$, i.e.,
$$
\Phi_m (m) = {\cal C}_{\rm n}m^{-2.5}, 
\eqno(3.2.1)
$$
where ${\cal C}_{\rm n}$ is the normalization constant.
From the surface mass density computed by 
$\Sigma_m = (\Sigma_L /L_{\rm bulge})M_{\rm pop}=
2814\ M_\odot\ {\rm pc}^{-2}$
the value of ${\cal C}_{\rm n}$ is obtained by
$$
{\cal C}_{\rm n}= {\Sigma_m \over \int \Phi'_mmdm},
\eqno(3.2.2)
$$ 
where $\Phi'_m$ is an arbitrarily normalized MF.
The model III MF is shown in the the upper panel of Figure 4 (solid line)
and it is compared with model II MF (dotted line).
Based on this MF
we reproduce the event rate distribution and it is shown in
Figure 4 (lower panel).
One finds that the event rate distribution matches impressively 
well with the observed one.

However, this simple picture does not hold up under closer examination.
First of all, all bulge mass cannot be in objects $m < 0.6\ M_\odot$,
since bulge MF has been measured for objects $m>0.6\ M_\odot$.
According to our MF of stars and WDs, these surely-existing relatively 
massive objects account for important fraction of total number and mass
of the bulge; $\sim 31\%$ of the total number and $\sim 51\%$ of 
the total mass of stars. 
Therefore, they should be included in optical depth and event rate 
computation.
To make up these massive stars, model III MF overestimates low-mass stars, 
whose MF also does not match with observation.

\acknowledgements
We would like to thank A. Gould for making precious comments and
suggestions.

\clearpage

\clearpage

\begin{center}
\bigskip
\bigskip
\centerline{\small {TABLE 1}}
\smallskip
\centerline{\small {\sc The Lens Population Inventory}}
\smallskip
\begin{tabular}{cccrrrr}
\hline
\hline
\multicolumn{1}{c}{model} &
\multicolumn{1}{c}{parameter} &
\multicolumn{1}{c}{unit} &
\multicolumn{4}{c}{populations} \\
\multicolumn{1}{c}{} &
\multicolumn{1}{c}{} &
\multicolumn{1}{c}{} &
%\multicolumn{1}{c}{obs.} &
\multicolumn{1}{c}{stars} &
\multicolumn{1}{c}{WDs} &
\multicolumn{1}{c}{BDs} &
\multicolumn{1}{c}{total} \\
\hline
I   & $m_{\rm pop}$ & $10^{10}\ M_\odot$ &1.17  &0.28 &     &  1.45  \\
    & $\Sigma_N$    & ${\rm objects}\ {\rm pc}^{-2}$ &3422  &614 &  & 4036  \\
    & $\Sigma_m$ & $M_\odot\ {\rm pc}^{-2}$ &1562 &369 &     &  1931  \\
    & $\Sigma_L$ & $L_\odot\ {\rm pc}^{-2}$ &2412 & &   &  2412  \\
\bigskip
    & $N/L$  &      &      &      &     &  1.66   \\

II  & $m_{\rm pop}$ & $10^{10}\ M_\odot$ &1.17  &0.28 &0.65  &  2.10  \\
    & $\Sigma_N$    & ${\rm objects}\ {\rm pc}^{-2}$ &3422 &614 &10887 &14923 \\
    & $\Sigma_m$ & $M_\odot\ {\rm pc}^{-2}$ &1562 &369 &871   &  2802  \\
    & $\Sigma_L$ & $L_\odot\ {\rm pc}^{-2}$ &2412 & &   &  2412  \\
\bigskip
    & $N/L$   &      &      &     &    &  5.33   \\

III & $m_{\rm pop}$ & $10^{10}\ M_\odot$ &2.10 &     &      & 2.10   \\
    & $\Sigma_N$    & ${\rm objects}\ {\rm pc}^{-2}$ & 16283 & &  & 16283  \\
    & $\Sigma_m$ & $M_\odot\ {\rm pc}^{-2}$ & 2814 &   &   & 2814   \\
    & $\Sigma_L$ & $L_\odot\ {\rm pc}^{-2}$ & 2412 &  &    & 2412  \\
    & $N/L$   &      &      &      &     & 6.75   \\
\hline
\end{tabular}
\end{center}
\smallskip
\noindent
{\footnotesize
\qquad NOTE.---
The lens population inventory of various mass function models.
In the table $m_{\rm pop}$ represents the total mass of each lens
population within the galactic bulge, and
$\Sigma_N$, $\Sigma_m$, and $\Sigma_L$ are the surface number, mass, and 
light density seen through BW.
All these surface densities are normalized for objects in an 
area of ${\rm pc}^2$ at the galactic center.
The number-to-light ratio is determined by
$N/L = \Sigma_N/\Sigma_L$.
}

\clearpage

\bigskip
\postscript{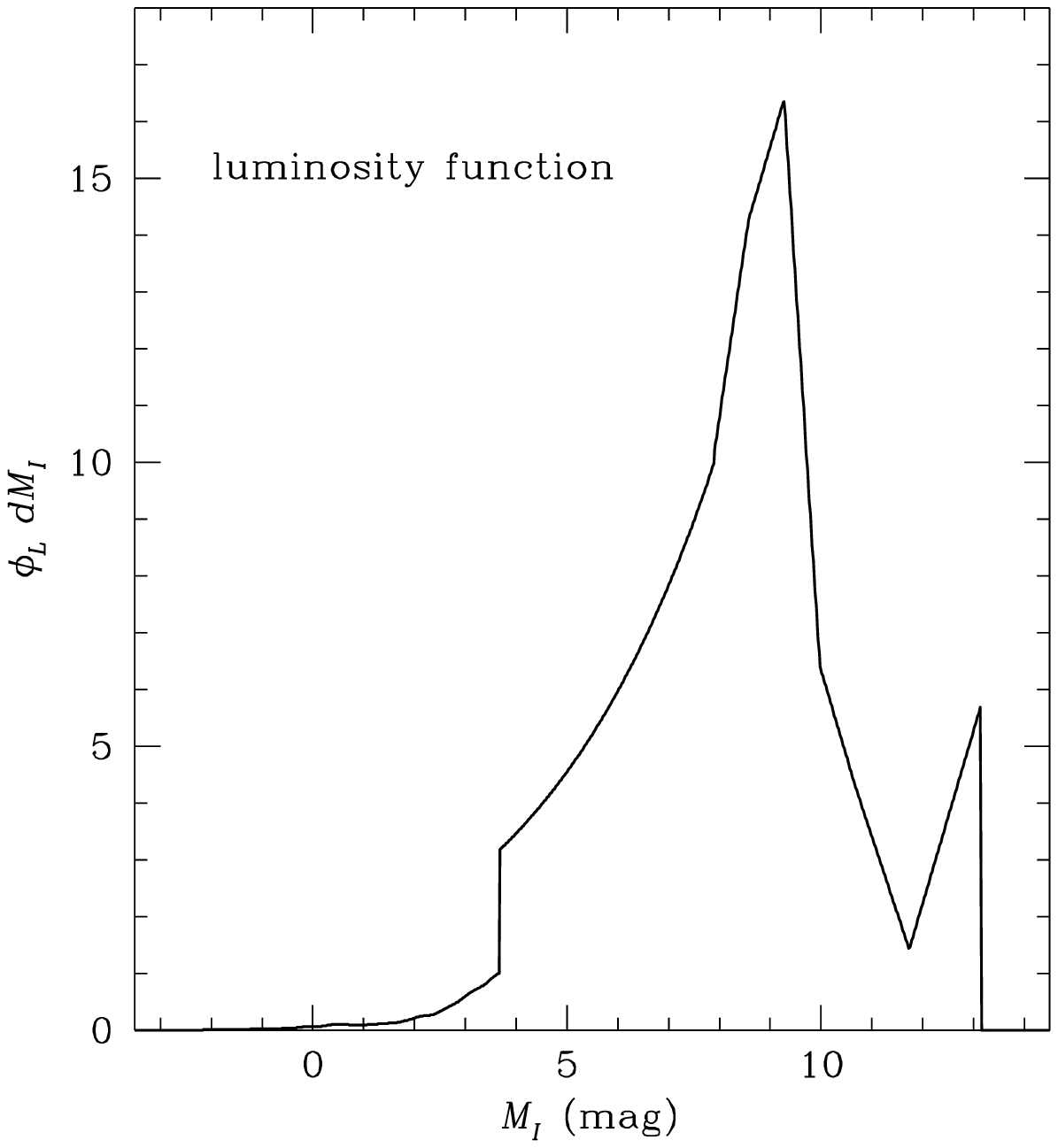}{0.72}
\noindent
{\footnotesize {\bf Figure 1:}\
The absolute $I$-band LF of galactic bulge stars.
The LF is normalized for stars in a physical area of 
${\rm pc}^2$ at the galactic center, corresponding to an angular 
area of $(0.43\ {\rm arcmin})^2$, and its range,
$-4 \leq M_I \leq 15$ is divided into 1000 intervals.
}

\clearpage

\bigskip
\postscript{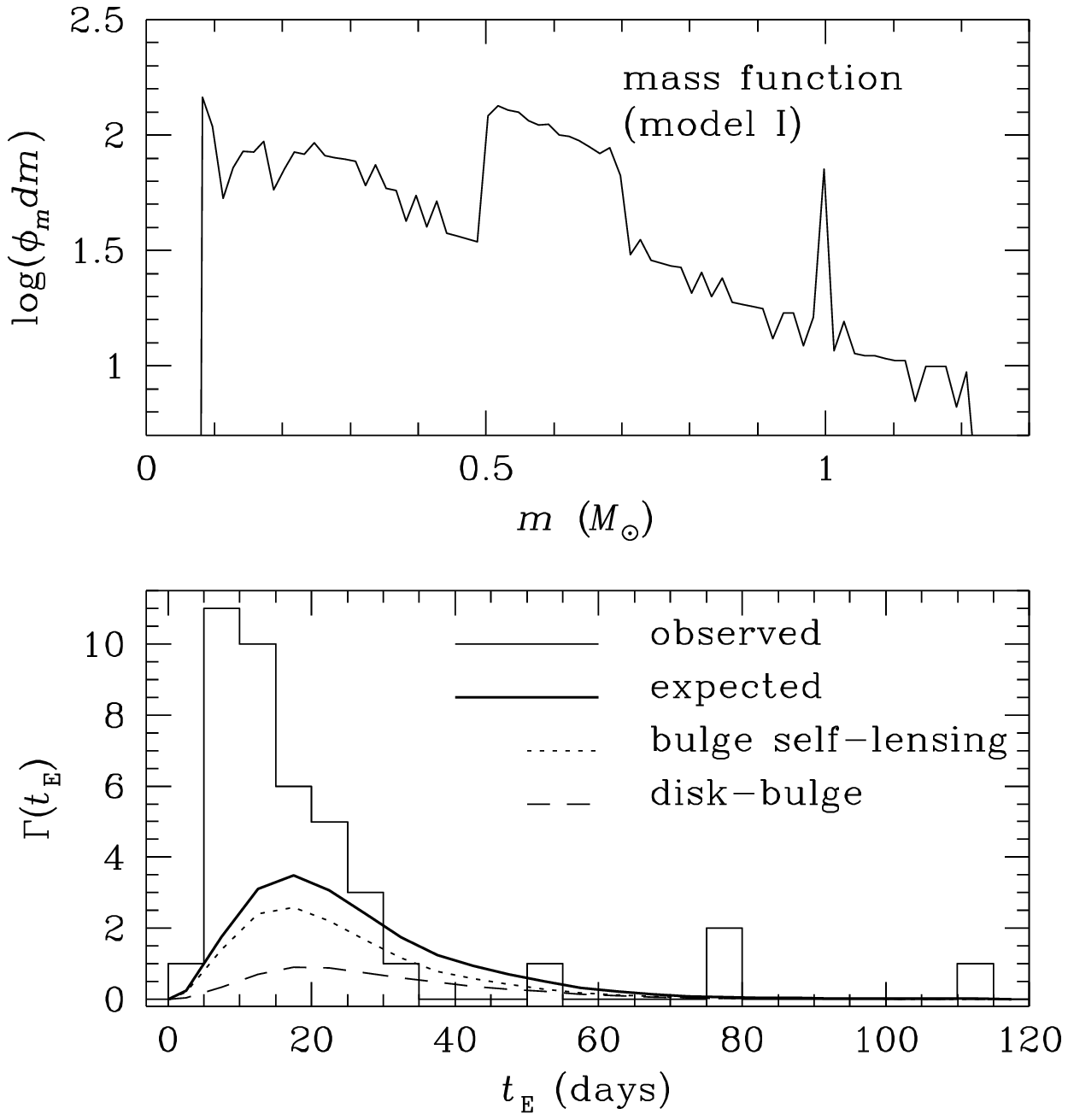}{0.72}
\noindent
{\footnotesize {\bf Figure 2:}\
The MF of known population of lenses, stars+WDs (upper panel),
and resulting event rate distribution (lower panel).
The MF is normalized for objects in an area of ${\rm pc}^2$
at the galactic center, and its
range, $0\ M_\odot \leq m \leq 1.5\ M_\odot$, is
divided into 100 intervals.
The total expected event rate distribution (thick solid line), 
which is sum of bulge-bulge (short-dashed line) and disk-bulge
(long dashed line),  
is compared with that of MACHO experiment (histogram).
It is clear that the event rate distribution from the known population of 
stars and WDs alone cannot explain the observed distribution.
They make up, respectively, just $\sim 49\%$ and $\sim 37\%$ of observed 
event rate and optical depth.
Note the difference between the two distributions arises especially in short 
time scale region.
}

\clearpage

\bigskip
\postscript{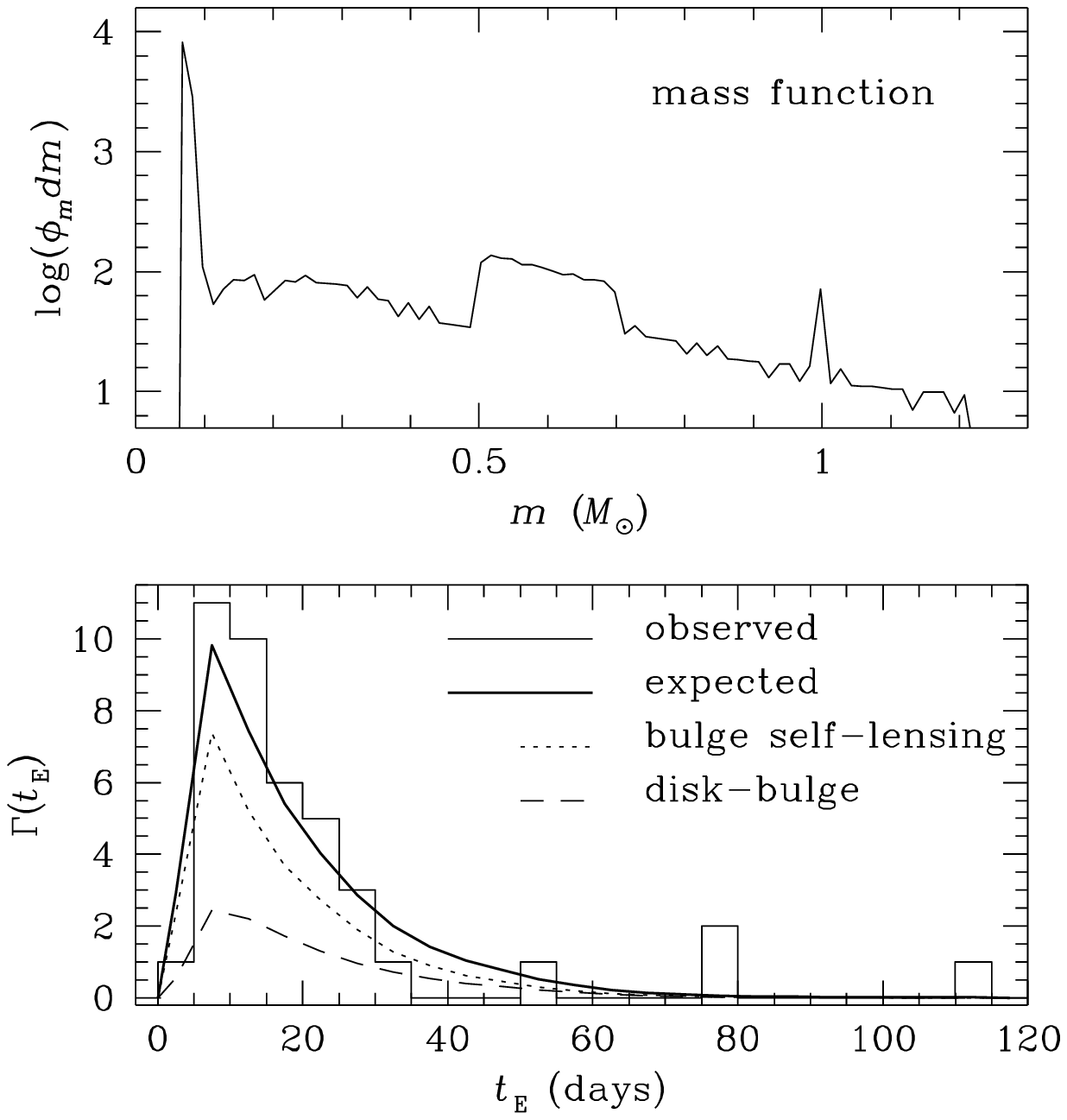}{0.72}
\noindent
{\footnotesize {\bf Figure 3:}\
Model II MF (upper panel) in which lenses are composed of known 
population of stars and WDs plus BDs, which make up the rest of 
dynamically determined galactic bulge mass.
The resulting event rate distribution is shown in the lower panel.
The notations are same as Figure 2.
One finds that the expected distribution matches very well
with the observed distribution.
}

\clearpage

\bigskip
\postscript{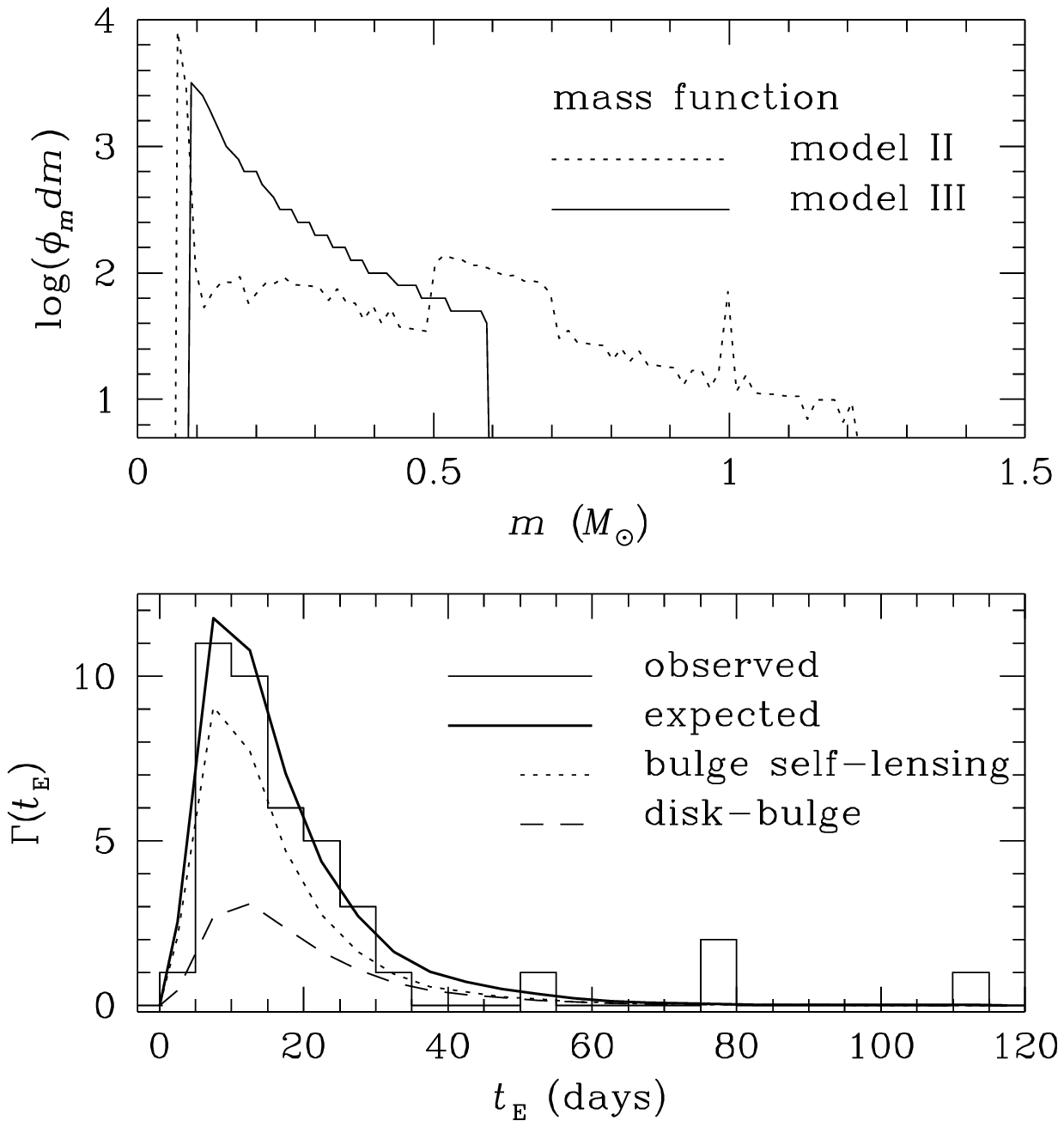}{0.72}
\noindent
{\footnotesize {\bf Figure 4:}\
Model III MF (upper panel) and corresponding event rate distribution 
(lower panel).
The model follows a Salpeter power-law MF in the mass range 
$0.08\ M_\odot \leq  m \leq 0.6\ M_\odot$. 
The notations are same as Figure 2.
Compared to model II MF, model III MF does not include
surely existing brighter, and thus massive, stars which make
up nearly half of the total mass of stars.
Despite this unrealistic model, it produces event rate distribution
that matches with the observed one (lower panel).
}

\end{document}